%% file: main.tex
\documentclass[a4paper,11pt]{ETHpaper}

\usetikzlibrary{positioning, calc}
\usetikzlibrary{decorations}
\usetikzlibrary{arrows.meta}
\usetikzlibrary{arrows}
\usetikzlibrary{decorations.shapes}

\usepackage{xcolor}
\usepackage{lipsum}

\definecolor{colorSG}{RGB}{168,50,45}
\definecolor{customY}{HTML}{FBB13C}
\definecolor{customG}{HTML}{218380}
\definecolor{customT}{HTML}{73D2DE}
\definecolor{customW}{HTML}{FCFCFF}
\definecolor{customB}{HTML}{2E5EAA}
\definecolor{customP}{HTML}{5B4E77}
\definecolor{gitRedFaded}{HTML}{FFEEF0}
\definecolor{gitGreenFaded}{HTML}{E6FFED}
\definecolor{gitRed}{HTML}{FFDCE0}
\definecolor{gitGreen}{HTML}{CDFFD8}
\definecolor{gitRedFull}{HTML}{CB2431}
\definecolor{gitGreenFull}{HTML}{2CBE4E}

\makeatletter
\def\blfootnote{\gdef\@thefnmark{}\@footnotetext}
\makeatother

\begin{document}

\title{Predicting Influential Higher-Order Patterns in Temporal Network Data}

\titlealternative{Predicting Influential Higher-Order Patterns in Temporal Network Data}

\author{Christoph Gote\textsuperscript{1,2,4} \qquad Vincenzo Perri\textsuperscript{1,5} \qquad Ingo Scholtes\textsuperscript{1,3}}
\authoralternative{Christoph Gote, Vincenzo Perri, Ingo Scholtes}
\address{
\textsuperscript{1}Data Analytics Group, University of Zurich, Zurich, Switzerland \\
\textsuperscript{2}Chair of Systems Design, ETH Zurich, Zurich, Switzerland \\
\textsuperscript{3}Chair of Machine Learning for Complex Networks, University of Würzburg, Würzburg, Germany \\[2mm]
\textsuperscript{4}cgote@ethz.ch\\
\textsuperscript{5}perri@ifi.uzh.ch
}
\www{}
\reference{Ver. of: \today}
\maketitle

\blfootnote{Contributed equally.}

\begin{abstract}
    \noindent Networks are frequently used to model complex systems comprised of interacting elements.
    While edges capture the topology of \emph{direct} interactions, the true complexity of many systems originates from higher-order patterns in paths by which nodes can \emph{indirectly} influence each other.
    Path data, representing ordered sequences of consecutive direct interactions, can be used to model these patterns.
    On the one hand, to avoid overfitting, such models should only consider those higher-order patterns for which the data provide sufficient statistical evidence.
    On the other hand, we hypothesise that network models, which capture only direct interactions, underfit higher-order patterns present in data.
    Consequently, both approaches are likely to misidentify influential nodes in complex networks.
    We contribute to this issue by proposing five centrality measures based on MOGen, a multi-order generative model that accounts for all indirect influences up to a maximum distance but disregards influences at higher distances.
    We compare MOGen-based centralities to equivalent measures for network models and path data in a prediction experiment where we aim to identify influential nodes in out-of-sample data.
    Our results show strong evidence supporting our hypothesis.
    MOGen consistently outperforms both the network model and path-based prediction.
    We further show that the performance difference between MOGen and the path-based approach disappears if we have sufficient observations, confirming that the error is due to overfitting.
    \\[1em]
\end{abstract}

\newpage

\section{Introduction} \label{sec:introduction}

Network models have become an important foundation for the analysis of complex systems across various disciplines, including physics, computer science, biology, economics, and the social sciences \cite{strogatz2001exploring}.
To this end, we commonly utilise network models in which \emph{nodes} represent the interacting elements, and \emph{edges} represent dyadic interactions between those elements.
A significant contribution of this perspective on complex systems is that it provides a unified mathematical language to study how the topology of the interactions between individual elements influences the macroscopic structure of a system or the evolution of dynamical processes \cite{boccaletti2006complex}.

In a network, edges capture the \emph{direct} influence between adjacent nodes.
However, for most networked systems with sparse interaction topologies, the true complexity originates from higher-order patterns capturing \emph{indirect} influence mediated via \emph{paths}, i.e., via sequences of incident edges traversed by dynamical processes.
The general importance of paths for analysing complex systems is expressed in many standard techniques in social network analysis and graph theory.
Examples include measures for the importance of nodes based on shortest paths \cite{freeman1977set,bavelas1950communication}, methods for the detection of community structures that are based on paths generated by random walkers \cite{rosvall2008maps}, but also algebraic and spectral methods that are based on powers of adjacency matrices or the eigenvalues of graph Laplacians \cite{chung1997spectral}, which can be thought as implicitly expanding edges into paths.

Standard network methods typically analyse systems based on paths that are generated by some model or algorithm operating on the network topology, e.g., shortest paths calculated by an algorithm, random paths generated by a stochastic model, or all paths transitively expanded based on the network topology.
The choice of a suitable model or process generating those paths is a crucial step in network analysis, e.g., for the assessment of node importance \cite{borgatti2005centrality}.
On the other hand, rather than using paths generated by models, we often have access to time-series data that captures real paths in networked systems.
Examples include human behavioural data such as time-stamped social interactions, clickstreams on websites, or travel itineraries in transportation networks.

Recent works have shown that, for many complex systems, the patterns in time series data on such paths cannot be explained by the network topology alone.
They instead contain higher-order patterns that influence the causal topology of a system, i.e., who can indirectly influence whom over time.
To capture these patterns, higher-order generalisations of network models have been proposed \cite{battiston2020networks,torres2020and,lambiotte2019networks}.
While the specific assumptions about the type of higher-order structures included in those models differ, they have in common that they generalise network models towards representations that go beyond pairwise, dyadic interactions.
Recent works in this area have used higher-order models for non-Markovian patterns in paths on networks to study random walks and diffusion processes \cite{scholtes2014causality,rosvall2014memory,lambiotte2015effect}, detect communities and assess node centralities \cite{rosvall2014memory,scholtes2016higher,xu2016representing,edler2017mapping,peixoto2017modelling}, analyse memory effects in clinical time series data \cite{palla2018complex,krieg2020higher,myall2021network}, generate node embeddings and network visualisations based on temporal network data \cite{saebi2020honem,tao2017honvis,perri2020}, detect anomalies in time series data on networks \cite{saebi2020efficient,larock2020hypa}, or assess the controllability of networked systems \cite{zhang2017controllability}.
Moreover, recent works have shown the benefit of \emph{multi-order models} that combine multiple higher-order models, e.g., for the generalisation of PageRank to time series data \cite{scholtes2017network} or the prediction of paths \cite{gote2020predicting}.



This work extends this view by making the following contributions:
\begin{itemize}
	\item We consider five centrality measures for nodes in complex networks and generalise them to MOGen, a multi-order generative model for paths in complex networks \citep{gote2020predicting}. Those measures can be considered proxies for the influence of specific node sequences on dynamical processes like, e.g., epidemic spreading and information propagation.
	\item We show that the direct use of observed paths to calculate those centralities yields better predictions of influential nodes in time series data than a simpler network-based model if there is sufficient training data. At the same time, this approach introduces a substantial generalisation error for small data sets. This motivates the need for a modelling approach that balances between under- and overfitting.
	\item We develop a prediction technique based on a probabilistic graphical model that integrates Markov chain models of multiple higher orders. Unlike previous works that used multi-order models to model paths in networks, our framework explicitly models the start and end nodes of paths. We show that this explicit modelling of start/end probabilities is crucial to predict influential node sequences.
	\item Using five empirical data sets on variable-length paths in human clickstreams on the Web, passenger trajectories in transportation systems, and interaction sequences in time-stamped contact networks, we show that our approach provides superior prediction performance.
\end{itemize}

\section{Methods}\label{sec:methods}

In the following, we introduce our approach to predict influential nodes and higher-order patterns based on MOGen, a multi-order generative model for path data \citep{gote2020predicting}.

\subsection{Paths on Network Topologies}\label{sec:paths_on_network_topologies}

We mathematically define a \textit{network} as tuple $G = (V,E)$, where $V$ is a set of nodes and $E$ is a set of edges.
In the example of a public transport system, the individual stations are the nodes, and an edge exists between two nodes if there is a direct connection between the two stations.
Users of the system move from start to destinations following \textit{paths} that are restricted by the network topology.
A \textit{path} is defined as an ordered sequence $s = v_1 \rightarrow v_2 \rightarrow \dots \rightarrow v_{l_s}$ of nodes $v_i \in V$, where $l_s$ is the length of the path and nodes can appear more than once.
We refer to a set of paths constrained by the same network topology as path data set $P$.

While empirical paths can come from various sources, we can differentiate between two main types: (i) data directly recorded in the form of paths; (ii) paths extracted from data on temporal interactions, i.e., a temporal network.
Examples for the first case include clickstreams of users on the Web or data capturing passenger itineraries from public transportation systems.
The primary example of temporal data are records on human interactions, which are a common source for studying knowledge transfer or disease transmission.

A \textit{temporal network} is a tuple $G^{(t)}=(V,E^{(t)})$, where $V$ is a set of vertices and $E^{(t)}$ is a set of edges with a time stamp $E^{(t)} \subseteq V \times V \times \mathbb{N}$.
We can extract paths from a temporal network by setting two conditions. 
First, for two time edges $e_i = (v_1,v_2;t_1)$ and $e_j = (v_2,v_3;t_2)$ to be considered consecutive in a path---i.e., $s = \cdots \rightarrow v_1 \rightarrow v_2 \rightarrow v_3 \rightarrow \cdots$---they have to respect the arrow of time, i.e., $t_1 < t_2$.
Second, consecutive interactions belong to the same path only if they occur within a time window $\delta$, i.e., $t_2 - t_1 \leq \delta$.
Using these conditions, we can derive a set of paths $P$ from any temporal network.

In summary, the network topology constrains the paths that are possible in real-world systems, such as transport or communication systems. 
However, empirical path data contain additional information on the start and endpoints of paths and the sequences in which nodes are traversed that the network topology does not capture.

\subsection{Modelling Higher-Order Patterns in Path Data}\label{sec:modelling_higher_order_patterns}
In the previous section, we showed that empirical paths capture information not contained in the network topology. 
Based on our arguments, one might assume that paths are always better to capture the dynamics on a networked system compared to the topology alone.
However, the validity of this argument strongly depends on the number of paths that we have observed.

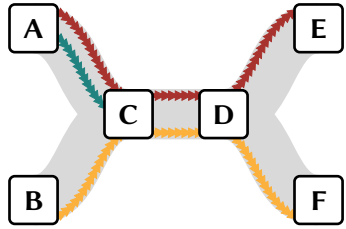
\begin{figure}
	\centering
	\begin{tikzpicture}\sffamily
		\tikzstyle{path}=[decorate,decoration={triangles, shape size=3.5pt ,segment length=2.5pt}, line width=0pt, fill=black]
		\coordinate (vd) at (.5,-1);
		\def\s{1.035}
		
		\coordinate (A) at ($(-1,0)+(vd)+\s*(0,1.1)$);
		\coordinate (B) at ($(-1,0)+(vd)+\s*(0,-1.1)$);
		\coordinate (C) at ($(.25,0)+(vd)+\s*(0,0)$);
		\coordinate (D) at ($(1.5,0)+(vd)+\s*(0,0)$);
		\coordinate (E) at ($(2.75,0)+(vd)+\s*(0,1.1)$);
		\coordinate (F) at ($(2.75,0)+(vd)+\s*(0,-1.1)$);
		
		\coordinate (h) at (3.5pt,0);
		\coordinate (v) at (0,3.5pt);
		
		\coordinate (h2) at (2pt,0);
		\coordinate (v2) at (0,2pt);
		
		\foreach \x/\y in {A/C,B/C,C/D,D/E,D/F}{
			\draw[line width=19pt, black!15] (\x) to[out=0, in=180] (\y);
		}

		\draw[path, colorSG] ($(A)+2*(v)+2*(h)$) to[out=-20, in=200] ($(C)+2*(v)+2*(h)$);
		\draw[path, colorSG] ($(C)+2*(v)$) -- ($(D)+2*(v)$);
		\draw[path, colorSG] ($(D)+2*(v)-2*(h)$) to[out=-20, in=200] ($(E)+2*(v)-2*(h)$);
		
		\draw[path, customG] ($(A)+0*(v)+0*(h)$) to[out=0, in=180] ($(C)+0*(v)+0*(h)$);
		
		\draw[path, customY] ($(B)-2*(v)+2*(h)$) to[out=20, in=160] ($(C)-2*(v)+2*(h)$);
		\draw[path, customY] ($(C)-2*(v)$) -- ($(D)-2*(v)$);
		\draw[path, customY] ($(D)-2*(v)-2*(h)$) to[out=20, in=160] ($(F)-2*(v)-2*(h)$);
		
		\foreach \Point in {A,B,C,D,E,F}{
			\draw [line width=1pt, rounded corners=2pt, fill=white] ($(\Point)+4.5*(h2)-4.5*(v2)$) rectangle node {\textbf{\Point}} ($(\Point)-4.5*(h2)+4.5*(v2)$);
		}
	\end{tikzpicture}
	\caption{Exemplary set of paths on a network topology. We observe three colour coded paths from $A$ to $B$ (\textcolor{customG}{\faSquare}), from $A$ to $E$ (\textcolor{colorSG}{\faSquare}), and from $B$ to $F$ (\textcolor{customY}{\faSquare}). The underlying network topology is shown in grey (\textcolor{black!15}{\faSquare})}
	\label{fig:toy_example}
\end{figure}

Let us consider the example shown in \Cref{fig:toy_example}.
As we can infer from the colour coded paths, a path in $D$ will always continue to $E$ if it started in $A$.
In contrast, if the path started in $B$, it will continue to $F$.
But does this mean that paths from $A$ to $F$ do not exist, despite being possible according to the underlying network topology?
To address this question, we need to consider how often we observed the paths from $A$ to $E$ and $B$ to $F$.
If, e.g., we observed both paths only once each, we would have little evidence suggesting that a path from $A$ to $F$ would not be possible.
Hence, in this case, using the observed paths as indicators for all possible paths would overfit the data, and a network model would be more appropriate.
In contrast, observing both paths many times without ever observing paths from $A$ to $F$ would indicate that paths from $A$ to $F$ do not exist or are at least significantly less likely than the observed paths. 
In this case, a network model would underfit the data by not adequately accounting for the patterns present in the empirical path data.

These examples underline that to capture the influence of nodes in real-world networked systems, neither a network model nor a limited set of observed paths is sufficient.
Instead, we require a model that can both represent the non-Markovian patterns in the path data, and allow transitions that are consistent with the network topology and cannot be ruled out because path data have not provided enough evidence.

\subsection{MOGen}

Our work is based on MOGen, a multi-order generative model for paths~\citep{gote2020predicting} that combines information from multiple higher-order models.
In addition, MOGen explicitly considers the start- and end-points of paths using the special initial and terminal states $*$ and $\dagger$.
MOGen represents a path $v_1 \rightarrow v_2 \rightarrow \dots \rightarrow v_{l}$ as
\begin{align}
	* \rightarrow v_1 \rightarrow (v_1, v_2) \rightarrow \dots\rightarrow (v_{l-K+1}, \ldots, v_{l}) \rightarrow \dagger,
	\label{eq:path_representation}
\end{align}
where $K$ denotes the maximum memory the model accounts for.
Combining the representations of all paths in a set $P$, the resulting MOGen model is fully described by a multi-order transition matrix $\textbf{T}^{(k)}$ shown in \Cref{fig:MOGen_representation}.
The entries $\textbf{T}^{(k)}_{ij}$ of $\textbf{T}^{(k)}$ capture the probability of a transition between two higher-order nodes.

Considering no memory, a MOGen model with $K=1$ is equivalent to a network model but for nodes $*$ and $\dagger$ that additionally consider starts and ends of paths.
In turn, a MOGen model with $K$ matching the maximum path length observed in $P$ is a lossless representation of the set of paths.
Thus, MOGen allows us to find a balance between the network model---allowing all observed transitions in any order---and the observed set of paths---only allowing for transitions in the order in which they were observed.

\paragraph{MOGen: Fundamental matrix}

\input{MOGen_representation}

Building on the original model \citep{gote2020predicting}, we interpret the multi-order transition matrix $\textbf{T}^{(K)}$ of MOGen as an absorbing Markov chain where the states $(v_1, \ldots, v_{n-1}, v_n)$ represent a path in node $v_n$ having previously traversed nodes $v_1, \ldots, v_{n-1}$.
Using this interpretation allows us to split $\textbf{T}^{(K)}$ into a transient part $\textbf{Q}$ representing the transitions to different nodes on the paths and an absorbing part $\textbf{R}$ describing the transitions to the end state $\dagger$.
We can further extract the starting distribution $\textbf{S}$.
All properties are represented in \Cref{fig:MOGen_representation}.

This representation allows us to compute the Fundamental matrix $\textbf{F}$ of the corresponding Markov chain.
\begin{align}
	\textbf{F} = \left(\textbf{I}^{(n\times n)} - \textbf{Q}\right)^{-1}
\end{align}
Here, $\textbf{I}^{(n\times n)}$ is the $n \times n$ identity matrix, where $n$ is the number of nodes in the multi-order model without counting the special states $*$ and $\dagger$.
Entries $(i, j)$ of this Fundamental matrix $\textbf{F}$ represent the expected number of times a path in node $i$ will visit node $j$ before ending.
The Fundamental matrix $\textbf{F}$ is essential as it allows us to compute path centrality measures for the MOGen model \emph{analytically}.

\subsection{Centrality measures}\label{sec:centrality_measures}

We now introduce five MOGen-based centrality measures that we use in our comparison.
For all MOGen-based centrality measures, we also introduce the corresponding measures for the network and a set of paths.

\subsubsection{Betweenness Centrality}
Betweenness centrality considers nodes as highly important if they frequently occur on paths connecting pairs of other nodes.
In a network, the betweenness centrality of a node $v$ is given by the ratio of shortest paths $\sigma_{st}(v)$ from $s$ to $t$ through $v$ to all shortest paths from $s$ to $t$ $\sigma_{st}$ for all pairs of nodes $s$ and $t$:
\begin{align}
	b_v^{(N)} = \sum_{}\frac{\sigma_{st}(v)}{\sigma_{st}}.
\end{align}

Standard betweenness centrality calculated in a network model relies on the assumption that only shortest paths are used to connect two nodes.
Using actual path data, we can drop this assumption and consider paths that are \emph{actually} used.
Therefore, we can obtain the betweenness of a node in a given set of paths $P$ by simply counting how many times a node appears between the first and last node of all paths.

For MOGen, we can utilise the properties of the Fundamental matrix $\textbf{F}$.
Entries $(v,w)$ of $\textbf{F}$ represent the number of times we expect to observe a node $w$ on a path continuing from $v$ before the path ends.
Hence, by multiplying $\textbf{F}$ with the starting distribution $\textbf{S}$, we obtain a vector containing the expected number of visits to a node on any path.
To match the notions of betweenness for networks and paths, we subtract the start and end probabilities of all nodes yielding
\begin{align}
	b_v^{(M)} = \left(\textbf{S}\cdot\textbf{F}\right)_v - s_v - e_v^{(M)}.\label{eq:betweenness}
\end{align}
\Cref{eq:betweenness} allows us to compute the betweenness centrality for all nodes in the MOGen model---i.e. higher-order nodes.
The betweenness centrality of a first-order node $v$ can be obtained as the sum of the higher-order nodes ending in $v$.

\subsubsection{Closeness Centrality (Harmonic)}
When considering the closeness centrality of a node $v$, we aim to capture how easily node $v$ can be reached by other nodes in the network.
For networks, we are therefore interested in a function of the distance of all nodes to the target node $v$.
The distance matrix $\textbf{D}$ capturing the shortest distances between all pairs of nodes can be obtained, e.g., by taking powers of the binary adjacency matrix of the network where the entries at the power $l$ represent the existence of at least one path of length $l$ between two nodes.
This computation can be significantly sped up by using graph search algorithms such as the Floyd-Warshall algorithm \citep{floyd1962algorithm} used in our implementation.
As our networks are based on path data, the resulting network topologies are directed and not necessarily connected.
We, therefore, adopt the definition of closeness centrality for unconnected graphs, also referred to as harmonic centrality \citep{marchiori2000harmony}.
This allows us to compute the closeness centrality of a node $v$ as
\begin{align}
	c_v^{(M)} = \sum_{d \in \textbf{D}_v}\frac{1}{d},\label{eq:closeness_centrality}
\end{align}
where $\textbf{D}_v$ is the $v$-th row of $\textbf{D}$.

As MOGen models contain different higher-order nodes, $\textbf{D}$ captures the distances between higher-order nodes based on the multi-order network topology considering correlations up to length $K$.
While we aim to maintain the network constraints set by the multi-order topology, we are interested in computing the closeness centralities for first-order nodes.
We can achieve this by projecting the distance matrix to its first-order form, containing the distances between any pair of first-order nodes but constrained by the multi-order topology.
For example, for the distances $d\{(A, B), (C, A)\} = 3$ and $d\{(B, B), (C, A)\} = 2$, the distance between the first-order nodes $B$ and $A$ is $2$.
Hence, while for the network, the distances are computed based on the shortest path assumption, multi-order models with increasing maximum order $K$ allow us to capture the tendency of actual paths to deviate from this shortest path.
Based on the resulting distance matrix $\textbf{D}$, closeness centrality can be computed following \Cref{eq:closeness_centrality}.

Finally, for paths, the distance between two nodes $v$ and $w$ can be obtained from the length of the shortest sub-path starting in $v$ and ending in $w$ among all given paths.
Again, the closeness centrality is then computed using \Cref{eq:closeness_centrality}.
Therefore, while for all representations, we compute the closeness centrality of a node using the same formula, the differences in the results originate from the constraints in the topologies considered when obtaining the distance matrix $\textbf{D}$.

\subsubsection{Path End Probability}
The path end probability $e_v$ of a node $v$ describes the probability of a path to end in node $v$.
For paths, $e_v^{(E)}$ is computed correspondingly by counting the fraction of paths ending in node $v$.
For MOGen, all paths end with the state $\dagger$.
Therefore, $e_v^{(M)}$ is obtained from the transition probabilities to $\dagger$ of a single path starting in $*$.
This last transition can---and is likely to---be made from a higher-order node.
We can obtain the end probability for a first-order node by summing the end probabilities of all corresponding higher-order nodes.
The path end probability cannot be computed for a network model as the information on the start and end of paths is dropped for this representation.

\subsubsection{Path Continuation Probability}
When following the transitions on a path, at each point, the path can either continue or end.
With the path continuation probability $f_v$, we capture the likelihood of the path to continue from node $v$.
Similarly to the path start and end probabilities, we obtain the path continuation probability from a set of paths $P$ by counting the fraction of times $v$ does not appear as the last node on a path compared to all occurrences of $v$.

For MOGen, the path continuation probability is given directly by summing the probabilities of all transitions in the row of \textbf{$\textbf{T}^{(K)}$} corresponding to node $v$ leading to the terminal state $\dagger$.
As for other measures, for MOGen, the continuation probabilities are computed for higher-order nodes.
We can obtain continuation probabilities for a first-order node $v$ as the weighted average of the continuation probabilities of the corresponding higher-order nodes, where weights are assigned based on the relative visitation probabilities of the higher-order nodes.
As path information is required, no comparable measure exists for networks.

\subsubsection{Path Reach}
Finally, we consider path reach.
With path reach, we capture how many more transitions we expect to observe on a path currently in node $v$ before it ends.
To compute path reach for a set of paths $P$, we count the average number of nodes on all paths before the path ends for all nodes, in a procedure very similar to the one used to compute path closeness.
For MOGen, we can again use the properties of the Fundamental matrix $\textbf{F}$ and obtain the expected number as the row sum
\begin{align}
	\rho_v^{(M)} = \sum\textbf{F}_v - 1
\end{align}
We subtract 1 to discount for the occurrence of node $v$ at the start of the remaining path.
Analogous to the continuation probability, we obtain the path reach of a first-order node $v$ by weighting the path reach of all corresponding higher-order nodes according to their respective relative visitation probabilities.
Again, the path reach requires information on path ends.
Therefore, it cannot be computed using the network model.

\section{Analysis approach}

In \Cref{sec:methods}, we argued that network models are likely to \textit{underfit} patterns in observed paths that are due to some paths occurring less often (or not at all) while others appear more often than we would expect based on the network topology alone.
Similarly, we expect the centralities computed directly on the paths to \textit{overfit} these patterns.
We, therefore, expect that when computing centralities based on the network or the paths directly, we misidentify the nodes that are actually influential.
We further conjecture that the errors caused by overfitting are particularly severe if the number of observed paths is low, i.e., if we have insufficient data to capture the real indirect influences present in the complex system.

\begin{figure}
	\centering
	\includegraphics[width=\linewidth]{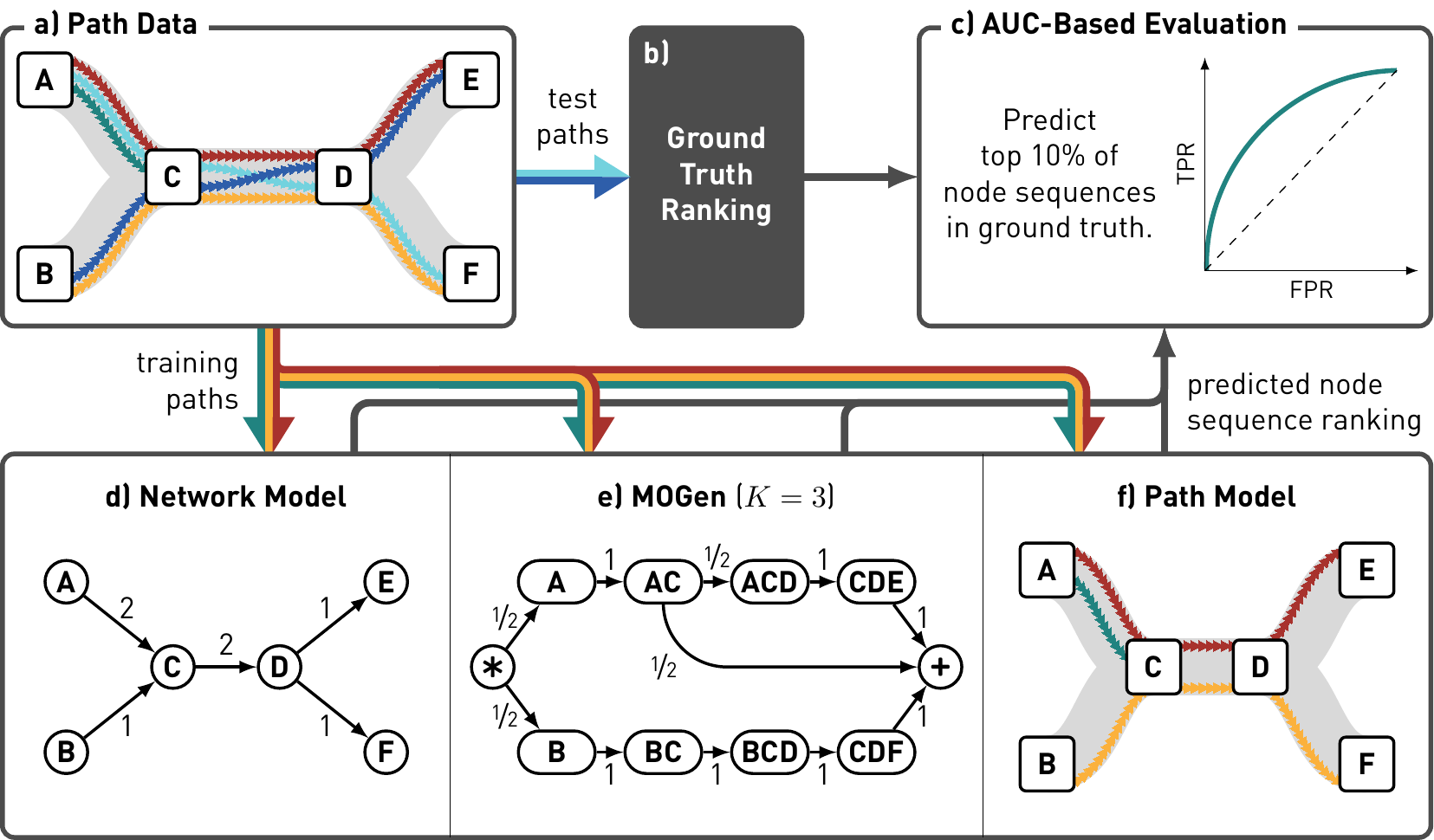}
	\caption{Overview of our approach to predict influential nodes and node sequences based on path data. We start from path data which we split into training and test sets. We learn three different models bases on the training data: (i) a network model containing all transitions from the training data, (ii) a multi-order generative model containing observed higher-order transitions up to a maximum order of $K$, which is determined by model selection, and (iii) a path model containing the full paths in the training set. Based on these models, we predict the influence of node or node sequences according to a broad range of centrality measures. We compare the ranking of node sequences to the ground truth rankings obtained from the test paths using AUC-based evaluation.}
	\label{fig:infographic}
\end{figure}

We now test our MOGen-based centrality against network- and path-based measures in five empirical path data sets.
To this end, we compare three types of models for a set of observed paths.
First, a network model containing all nodes and edges observed in the set of paths.
Second, a path model which precisely captures the observed paths, i.e., the model is identical to the set of paths.
Third, MOGen models with different maximum orders $K$ that capture all higher-order patterns up to a distance of $K$.

We operationalise our comparison in a prediction experiment in which we aim to predict influential nodes and higher-order patterns in a set of test data based on training data.
\Cref{fig:infographic} provides an overview of our evaluation approach.

\paragraph{Train-test split}
For our prediction experiment, we first split a given set of $N$ paths into a training and test set, while treating all observed paths as independent.
We denote the relative sizes of the training and test sets as $\nicefrac{n_{\text{tr}}}{N}$ and $\nicefrac{n_{\text{te}}}{N}$, respectively.

\paragraph{Ground truth ranking}
As introduced in \Cref{sec:methods}, our path-based centrality measures exclusively capture the importance of nodes in a set of observed paths.
While we expect this to lead to overfitting when making predictions based on training data, they yield precise ground truth influences when applied to the test data directly.
To obtain a ground truth ranking (see \Cref{fig:infographic}b), we sort the nodes and node sequences according to their influence in descending order.

\paragraph{Prediction of Influential Nodes and Node Sequences} 

The network model is the least restrictive model for a set of paths.
In contrast, the path model always considers the entire history.
With $K=1$, a MOGen model resembles a network model with added states capturing the start- and endpoints of paths.
By setting $K=l_{max}$, where $l_{max}$ is the maximum path length in a given set of paths, we obtain a lossless representation of the path data. 
By varying $K$ between $1$ and $l_{max}$, we can adjust the model's restrictiveness between the levels of the network and the path model.
We hypothesise that network and path models under- and overfit the higher-order patterns in the data, respectively, leading them to misidentify influential nodes and node sequences in out-of-sample data.
Consequently, by computing node centralities based on the MOGen model, we can reduce this error.

To test this, we train a network model, a path model, and MOGen models with $1 \leq K \leq 5$ to our set of training paths.
We then apply the centrality measures introduced in \Cref{sec:centrality_measures} to compute a ranking of nodes and node sequences according to each of the models.
In a final step, we compare the computed rankings to the ground truth ranking that we computed for our test paths.

\paragraph{Comparison to ground truth}

While our models are all based on the same set of training paths, they make predictions for node sequences up to different lengths.
We allow the comparison of different models' predictions through an upwards projection of lower-order nodes to their matching node sequences.
To this end, we match the prediction of the closest matching lower-order node $v_l \in \mathcal{L}$ as the prediction of the higher-order node $v_h \in \mathcal{H}$.
Here, $\mathcal{L}$ is the set of lower-order nodes, e.g., from the network model, whereas $\mathcal{H}$ is the set of higher-order nodes from the ground truth.
We define the closest matching lower-order node $v_l$ as the node with highest order in $\mathcal{L}$ such that $v_l$ is a suffix of $v_h$.

We evaluate how well the predictions match the ground truth using an AUC-based evaluation approach.
Our approach is built on a scenario in which we aim to predict the top $10\%$ most influential nodes and node sequences in the ground truth data.
By considering this scenario, we transform the comparison of rankings into a binary classification problem, where for each node or node sequence, we predict if it belongs into the top $10\%$ of the ground truth or not.
All results reported throughout this manuscript refer to averages over at least five validation experiments.

\paragraph{Datasets}

\input{dataset_description.tex}
\begin{figure*}[t!]
	\centering
	\includegraphics[width=\textwidth]{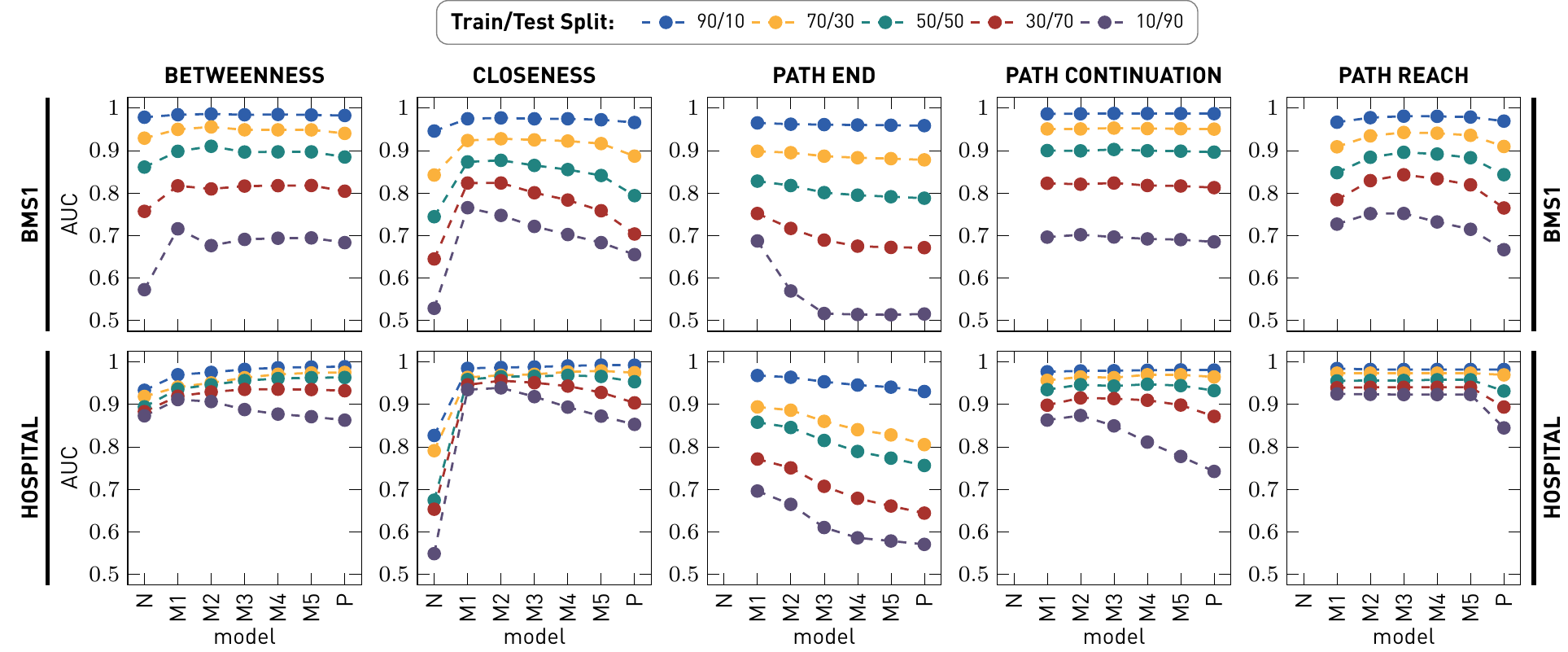}
	\caption{Prediction results for five centrality measures for the BMS1 and SCHOOL data sets and different train/test splits. N and P indicate the network and path model, respectively. M1 through M5 are MOGen models with maximum orders between 1 and 5.}
	\label{fig:results_network_measures}
\end{figure*}

We test our hypothesis in five empirical path data sets containing observations from three different categories of systems: (i) user clickstreams on the Web (BMS1~\cite{gazelle}), (ii) travel itineraries of passengers in a transportation network (TUBE~\cite{LTdata}), and (iii) time-stamped data on social interactions (HOSPITAL~\cite{Vanhems2013}, WORKPLACE~\cite{Genois2015_FaceToFace}, SCHOOL~\cite{stehle2011high}).
BMS1 and TUBE are directly collected in the form of paths.
For SCHOOL, HOSPITAL, and WORKPLACE we extracted paths following \Cref{sec:paths_on_network_topologies}, using $\delta$ as 800s, 1,200s, and 3,600s, respectively.
The raw data for all data sets are freely available online.
We provide summary statistics for all data sets in \Cref{tab:dataset_statistics}.

\section{Results}
\label{sec:results}

We now present the results of our prediction experiments comparing the performance of network, path, and MOGen models to predict the influence of nodes and node sequences in out-of-sample data.
For ease of discussion, we start our analysis focusing on the two data sets BMS1 and HOSPITAL.
\Cref{fig:results_network_measures} shows the results for our five centrality measures.
For betweenness and closeness, we do not require information on the start- and endpoint of paths.
Therefore, equivalent measures for the network model exist.
In contrast, no equivalent measures for the network model can be computed for path end, path continuation, and path reach.

We show the AUC values for the different models and for different relative sizes for our training and test sets.
The models shown on the $x$-axis are sorted according to the maximum distance at which they can capture indirect influences.
Thus, starting from the network model (N), via the MOGen models (M$K$) with increasing $K$, the models become more restrictive until ending with the path model (P).

Overall, the MOGen models outperform both the network model and the path models.
With less training data, the AUC scores of all models decrease.
However, as expected, these decreases are larger for the network and path models.
For the betweenness and closeness measures, this results in AUC curves that resemble ``inverted U-shapes''.
For the remaining measures, for which no equivalent network measures are available, we generally find that MOGen models with $K$ between 1 and 3 perform best and the prediction performance decreases for more restrictive models, such as the path model.
Our results highlight the risk of underfitting for network models and overfitting for path models. 
We further show that this risk increases when less training data is available.

\begin{table*}[t!]
	\caption{AUC values for all models and measures on five data sets for a 30/70 train-test split. N and P indicate the network and path model, respectively. M1 through M8 are MOGen models with maximum orders between 1 and 8 (shown in \textcolor{customG!20}{\faSquare}). The best performing result for each data set and measure is highlighted in bold.}
	\label{tab:full_result_table}
	\centering
	\includegraphics[width=\textwidth]{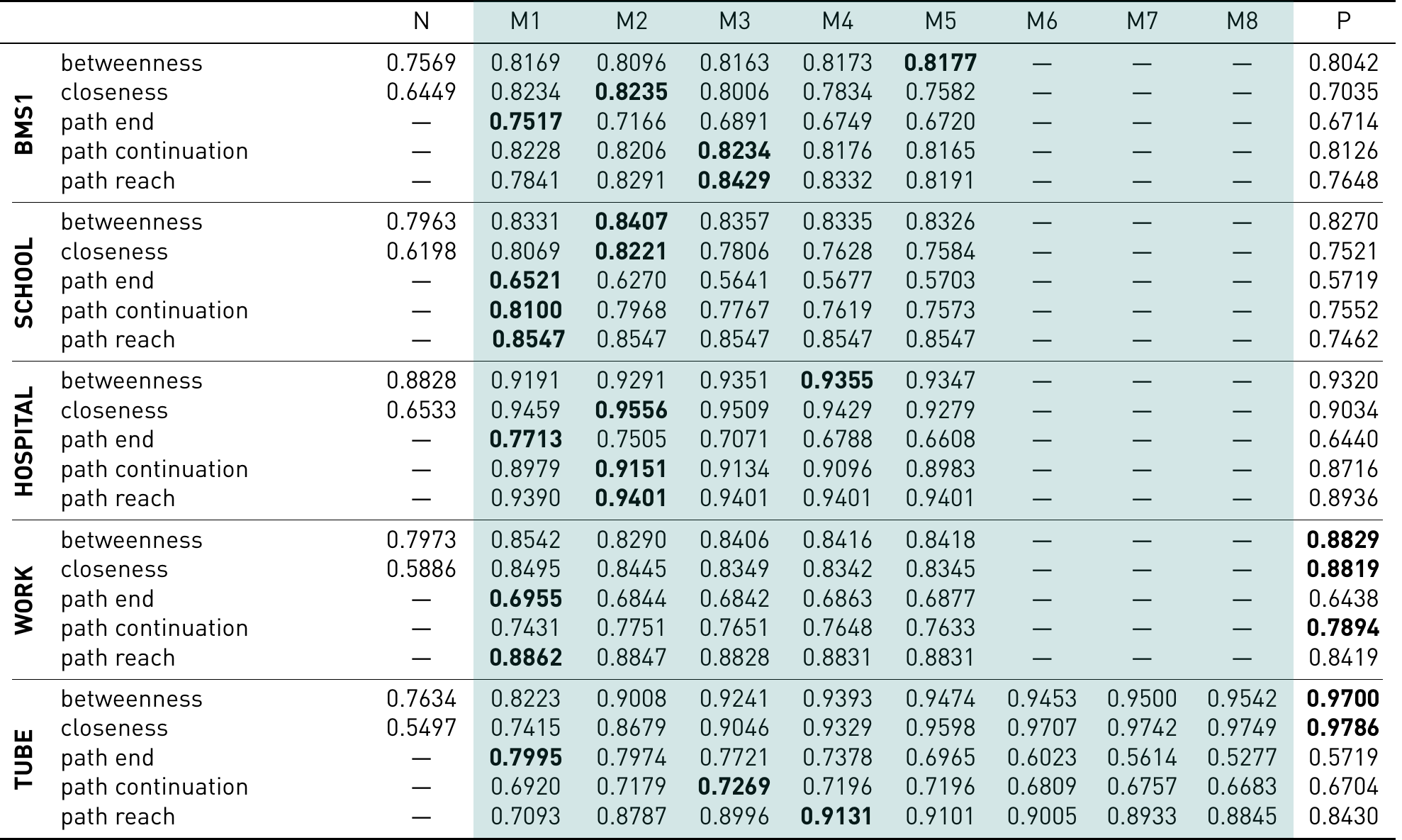}
\end{table*}

In \Cref{tab:full_result_table}, we show the results for all data sets and centrality measures for a 30/70 train/test split.
In general, we find similar patterns to those discussed with \Cref{fig:results_network_measures}.
However, for WORK and TUBE, the difference in prediction quality between the MOGen and path models decreases and for some measures, the path model even yields better performance.
WORK and TUBE are those data sets for which we have the highest fraction of total observed paths compared to the number of unique paths in the data sets.
As shown in \Cref{tab:dataset_statistics} BMS1 contains 59,601 total paths of which 18,473 are unique.
This means that, on average, each unique path is observed 3.2 times.
These counts increase to 4 for SCHOOL, 4.6 for HOSPITAL, 6.7 for WORK, and 132.9 for TUBE.
The good performance of the path model for these data sets shows that the error we found with fewer observations is indeed due to overfitting.
In other words, if we have a sufficient number of observations, we can compute the centralities on the path data directly.
However, if the number of observations is insufficient, the path model overfits the patterns in the training data and consequently performs worse on out-of-sample data.
How many observations are required to justify using the path model depends on the number of unique paths contained in the data set.

In conclusion, our results support our hypothesis.
By not capturing the higher-order patterns present in path data and not considering the start- and endpoints of paths, the network model consistently underfits the patterns present in path data.
Similarly, the path model overfits these patterns.
Consequently, when using either model to rank the influence of nodes and node sequences in path data, we obtain rankings that are not consistent with out-of-sample observations.
Prediction performance can be significantly improved by using MOGen models that prevent underfitting by capturing higher-order patterns up to a distance of $K$ while simultaneously preventing overfitting by ignoring patterns at larger distances.

\section{Conclusion}
\label{sec:conclusion}

Paths capture higher-order patterns, i.e., indirect influences, between elements of complex systems not captured by network topology.
To accurately capture the influence of nodes and node sequences, we must accurately account for these higher-order patterns present in our data.
However, not all higher-order patterns observed in a set of paths are representative of the actual dynamics of the underlying system.
In other words, by computing centralities on the full paths, we are likely to overfit higher-order patterns and attribute centrality scores to nodes and node sequences different to the ones we obtain when further observing the system and collecting additional paths.
Therefore, we require a model that captures only those higher-order patterns for which there is sufficient statistical evidence in the data.
We argued that the multi-order generative model MOGen is an ideal model for this purpose as it captures higher-order patterns in paths up to a given length while simultaneously including representations for the start and end of paths.

Based on the MOGen representation, we proposed measures to quantify the influence of both nodes and node sequences in path data according to five different notions of centrality.
Our centrality measures range from simple concepts like the betweenness to complex measures such as path reach.
For all centrality measures, we also proposed equivalent measures computed directly on path data.
While equivalent measures exist for the simple notions of centrality, networks cannot represent the start and end of paths and, hence, cannot represent the full information contained in a path.
Consequently, for the more complex measures, no network equivalents exist.

In a prediction experiment with five empirical data sets, we showed that networks models underfit and path models overfit higher-order patterns in path data.
Therefore, by computing the centralities of nodes or node sequences according to these models, we misidentify influential nodes.
By using MOGen, we can avoid both under- and overfitting.
Thus, when computing centralities for MOGen models, we obtain rankings that better represent influential nodes in out-of-sample data.

Our results highlight the potential consequences of applying networks---the most popular model for relational data---to sequential data. 
Similarly, MOGen-based centralities generally outperform those computed using the path model.
The performance difference is greater if the ratio between the number of observed paths and the number of unique paths in a data set decreases.
Thus, the larger the variance in the set of observed paths, the larger the potential for overfitting when using a path model to identify central nodes and node sequences in the data.
Large variances in observed paths characterise many real-world systems such as human interactions, where the range of possible interactions is extensive, and data is either costly to obtain or limited in availability.
In these cases, our MOGen-based centrality measures provide significantly more accurate predictions on the true influential nodes and node sequences compared to both the network- and path-based measures.

\section*{Archival and Reproducibility}
Sources for all data used in this paper are provided. A reproducibility package is available at \url{https://doi.org/10.5281/zenodo.7139438}.
A parallel implementation of the MOGen model is available at \url{https://github.com/pathpy/pathpy3}.

\section*{Acknowledgements}
All authors acknowledge support by the Swiss National Science Foundation, grant 176938.

\small

\bibliographystyle{sg-bibstyle}
\bibliography{bibliography}

\end{document}

%% file: MOGen_representation.tex
\begin{figure}
	\centering
	\begin{tikzpicture}[x=.7cm, y=-.7cm]
		\draw[very thick, rounded corners=2] (.5,.5) -| (.3,5) |- (.5, 9.5);
		\draw[very thick, rounded corners=2] (9.5,.5) -| (9.7,5) |- (9.5, 9.5);

		\draw[ultra thick, draw=white, fill=customY!30] (0.5,0.5) rectangle (9.5,9.5);
		\draw[ultra thick, draw=white, fill=black!10] (8.5,0.5) rectangle node {$0$} (9.5,1.5);
		\draw[ultra thick, draw=white, fill=customG!30] (2.5,.5) rectangle node {$0$} (8.5,1.5);
		
		\draw[ultra thick, draw=white, fill=customY!30] (2.5,1.5) rectangle node {\small$\textbf{T}_{1,2}$} (4.5,3.5);
		\draw[ultra thick, draw=white, fill=customY!30] (4.5,3.5) rectangle node[yshift=2] {\small$\vdots$} (5.5,5.5);
		\draw[ultra thick, draw=white, fill=customY!30] (5.5,5.5) rectangle node {\small$\textbf{T}_{K-1,K}$} (8.5,6.5);
		\draw[ultra thick, draw=white, fill=customY!30] (5.5,6.5) rectangle node {\small$\textbf{T}_{K,K}$} (8.5,9.5);
		\node at (3,6.5) {$0$};
		\node at (7,3.5) {$0$};
		
		\draw[ultra thick, draw=white, fill=customG!30] (.5,.5) rectangle node {\small$\textbf{T}_{0,1}$} (2.5,1.5);
		
		\draw[ultra thick, draw=white, fill=customB!30] (8.5,1.5) rectangle node {\small$\textbf{T}_{\dagger}$} (9.5,9.5);
		
		\node[anchor=south] at (1.5,0.5) {\small $V^1$};
		\node[anchor=south] at (3.5,0.5) {\small $V^2$};
		\node[anchor=south] at (5,0.3) {\small $\dots$};
		\node[anchor=south] at (7,0.5) {\small $V^K$};
		\node[anchor=south] at (9,0.5) {\small $\dagger$};
		
		\node at (-.8,1) {\small $*$};
		\node at (-.8,2.5) {\small $V^1$};
		\node at (-.8,4.3) {\small $\vdots$};
		\node at (-.8,6) {\small $V^{K-1}$};
		\node at (-.8,8) {\small $V^K$};
		
		\node[anchor=east] at (-2, 5) {$\textbf{T}^{(K)}=$};
	\end{tikzpicture}
	\caption{Multi-order transition matrix $\textbf{T}^{(K)}$ of a MOGen model with maximum-order $K$. We split $\textbf{T}^{(K)}$ into transient part $\textbf{Q}$ (\textcolor{customY!30}{\faSquare}) and absorbing part $\textbf{R}$ (\textcolor{customB!30}{\faSquare}). $\textbf{S}$ (\textcolor{customG!30}{\faSquare}) represents the starting distribution of paths.}\label{fig:MOGen_representation}
\end{figure}

%% file: dataset_description.tex
\begin{table}[tb!]
\caption{Summary statistics for our five empirical data sets.}\label{tab:dataset_statistics}

\footnotesize
\begin{tabularx}{\linewidth}{@{}X@{\hspace{-5mm}}rrrrrr@{}}
%

\toprule
{} & \multicolumn{2}{c}{paths} & \multicolumn{2}{c}{nodes on path} & \multicolumn{2}{c}{network topology} \\
\cmidrule(lr){2-3}\cmidrule(lr){4-5}\cmidrule(lr){6-7}
{} &  \multicolumn{1}{c}{total} &  \multicolumn{1}{c}{unique} &  \multicolumn{1}{c}{mean} & \multicolumn{1}{c}{median} & nodes & links \\
\midrule
BMS1  & 59,601 & 18,473 &     2.51 & 1 &            497 &          15,387 \\
TUBE & 4,295,731 & 32,313  &  7.9 & 7 &            276.0 &            663 \\
SCHOOL & 103,260 & 25,831 & 2.5 & 2 &            242 &           8,297 \\
HOSPITAL & 62,676 & 13,578 & 4.8 &            5 &             75 &           1,137 \\
WORK     & 7,832 & 1,170 &  2.5 & 2 &             92 &            753 \\
\bottomrule
\end{tabularx}
\end{table}